\documentclass[preprint,11pt]{elsarticle}

\usepackage{amssymb}
\usepackage{amsmath}
\usepackage{bm}
\usepackage{multirow}
\usepackage{xcolor}
\usepackage[normalem]{ulem}
\usepackage{soul}
\usepackage{arydshln}
\usepackage{subfig}

\begin{document}

\begin{frontmatter}

\title{Multiple-cell cavity design for high mass axion searches: an in-depth study}

\author[CAPP]{Junu Jeong}
%\ead{jwpc0120@ibs.re.kr}
\author[CAPP]{\corref{cor1}Sungwoo Youn}
%\ead{swyoun@ibs.re.kr}
\affiliation[CAPP]{organization={Center for Axion and Precision Physics Research, IBS},
            addressline={193, Munji-ro}, 
            city={Yuseong-gu},
            postcode={34051}, 
            state={Daejeon},
            country={Republic of Korea}}

\author[KHU]{Jihn E. Kim}

\affiliation[KHU]{organization={Department of Physics, Kyung Hee University},
            addressline={26, Kyunghee-daero},
            city={Dongdaemun-gu},
            postcode={02447},
            state={Seoul},
            country={Republic of Korea}}

\begin{abstract}
The invisible axion is a well-motivated hypothetical particle which could address two fundamental questions in modern physics -- the CP symmetry problem in the strong interactions and the dark matter mystery of our universe.
The plausible mass (frequency) range of the QCD axion as a dark matter candidate spans from $\mu$eV to meV ($\mathcal{O}$(GHz) to $\mathcal{O}$(THz)).
The axion haloscope using a resonant cavity has provided the most sensitive search method in the microwave region.
However, experimental searches have been limited to relatively low mass regions mainly due to the reduced cavity volume at high masses.
As an effective approach for high-mass axion searches, a unique cavity design, featured by multiple identical cells divided by equidistant thin metal partitions in a single cylindrical cavity, was proposed and successfully demonstrated.
We perform an in-depth study to characterize the multiple-cell cavity design and discuss the various advantages it offers for high-mass axion searches.
\end{abstract}

\begin{keyword}
Axion dark matter \sep Cavity haloscope \sep Multiple-cell
\PACS 14.80.Va \sep 95.35.+d
\end{keyword}

\end{frontmatter}

\section{\label{sec:intro}Introduction}

The axion is a hypothetical elementary particle induced by the Peccei-Quinn mechanism that was proposed to solve the CP problem in quantum chromodynamics (QCD)~\cite{PRL1977PQ,PRL1978Weinberg,PRL1978Wilczek}, one of the major fine tuning problems in particle physics.
Symmetry breaking at a sufficiently high energy scale could yield a very light axion which feebly couples to the standard particles.
Because the {\it invisible} axion would stably maintain a non-relativistic state, it has been considered as one of the strong candidates for cold dark matter~\cite{PLB1983Wilczek,PLB1983Abbott,PLB1983Dine}. 
The most sensitive method to search for the QCD axion dark matter in the microwave regime is the cavity haloscope which is designed to probe events of the axion-to-photon conversion inside a microwave cavity under a strong magnetic field~\cite{PRD1985Sikivie}.
The tiny signal of axion-induced photons are enhanced to an observable level by resonating with the cavity tuned to the axion mass. 

Since the mass of axion dark matter is unknown {\it a priori}, all possible masses should be probed by tuning the resonant frequency of the cavity.
How fast a given mass range can be scanned is quantified by the scan rate, which has the following proportionality in the limit of the cavity quality factor much smaller than the axion quality factor, i.e. $Q_{c} \ll Q_{a}$,~\cite{JCAP2020Kim}:
\begin{equation}
\label{eq:scan_rate}
    \frac{d\nu}{dt} \propto \left(\frac{g_{a\gamma\gamma}^{2}(\rho_{a}/m_{a})\langle \mathbf{B}_{0}^{2}\rangle V_{c} C}{k_{B} T} \right)^{2} Q_{c}Q_{a},
\end{equation}
where $g_{a\gamma\gamma}$ is the axion-photon coupling constant, $\rho_{a}$ and $m_{a}$ are the local axion density and its mass, $\langle \mathbf{B}_{0}^{2} \rangle$ is the square average of the external magnetic field over the cavity volume $V_{c}$, and $Q_{c}$ and $Q_{a}$ are the cavity and axion quality factors, respectively.
The form factor $C$, defined below, describes how well the electric field of the resonant mode ($\mathbf{E}_{r}$) is aligned with the applied magnetic field:
\begin{equation}
\label{eq:form_factor}
    C = \frac{|\int \mathbf{B}_{0}\cdot \mathbf{E}_{r}dV_{c}|^{2}}{\int |\mathbf{B}_{0}|^{2} dV_{c} \int \varepsilon |\mathbf{E}_{r}|^{2}dV_{c}},
\end{equation}
where $\varepsilon$ is the dielectric constant inside the cavity.
The cavity haloscope usually employs a solenoid magnet because it produces a fairly uniform and strong magnetic field in a local space.
Accordingly, the TM$_{010}$ mode of a cylindrical cavity is adopted for the resonant mode to take advantage of the largest form factor, yielding a large effective detection volume. 

The haloscope technique is almost the only experimental method that can reach sensitivity to the theoretical models of axion~\cite{PRL1979Kim,NPB1980SVZ,YF1980Zhitnitsky,PLB1981DFS} in the microwave regime.
However, there is a disadvantage that the sensitivity decreases rapidly as the search frequency increases.
It is mainly because the resonant frequency of the TM$_{010}$ mode is inversely proportional to the cavity radius, giving rise to substantial loss in the detection volume and gradual decrease in cavity quality factor.
In addition, the fundamental noise level of a quantum device, subject to the standard quantum limit, increases linearly with frequency.
Taking these effects into account, the scan rate decreases with the search frequency $\omega_{c}$ by
\begin{equation}
\label{eq:scan_rate_for_omega_c}
    \frac{d\nu}{dt} \propto \omega_{c}^{-20/3},
\end{equation}
which indicates for example that doubling the search frequency reduces the scanning speed by a factor of 100.

Various ideas have been proposed to address these issues.
One of the intuitive ways to recover the detection volume is to configure the system with multiple identical small cavities filling a given magnet bore~\cite{UCD2001Kinion,MCD2020Yang}.
In such a multiple-cavity system, a major challenge is to match the resonant frequency of the individual cavities to an acceptable level not to broaden the resonance width of the combined system~\cite{AP2018Jeong}.
Another approach exploits higher-order modes of cylindrical cavities with a precisely positioned hollow-shaped dielectric segments~\cite{JPG2020Kim,PRA2020Quiskamp}, which benefits from high $Q$ values but suffers from low frequency tunability.
Some other interesting proposals were made for searches in even higher mass regions by configuring a periodic structure of dielectric discs~\cite{PRL2017MADMAX} or metallic wires~\cite{PRL2019Lawson}.
However, handling a large number of base material may encounter engineering challenges in accurate alignment.

A new cavity concept, called multiple-cell cavity, was proposed by the Center for Axion and Precision Physics Research of the Institute for Basic Science (IBS-CAPP).
It was designed to minimize the volume loss for high-mass axion searches and to get rid of the necessity of frequency matching unavoidable in multiple-cavity systems~\cite{PLB2018Jeong}.
The effectiveness of the cavity design was successfully demonstrated by conducting an axion search experiment using a double-cell cavity, which set a new exclusion limit around 13.5~$\mu$eV~\cite{PRL2020Jeong}.
The result verified that the double-cell cavity boosts the scan rate by a factor of 4 compared to a conventional single cavity at the same mass.
Other experiments at CAPP adopted this new design with higher-cell multiplicities for efficient searches around 25\,$\mu$eV and 30\,$\mu$eV.

In this work, we report an in-depth study on the multiple-cell cavity design to generalize the characteristics as an effective haloscope for high-mass axion searches and to address the potential issues related to field localization.
Section~\ref{sec:pizza} explains the basic concept of the cavity design and discuss the various advantages over other ideas.
In Sec.~\ref{sec:chara}, major features of this new concept are described based on analytical calculations and numerical simulations.
Section~\ref{sec:local} provides a methodology for obtaining the optimal geometry to minimize the field localization owing to fabrication tolerance.
Finally, in Sec.~\ref{sec:discussion}, we briefly discuss the impact of this cavity design on axion search experiments in high mass regions.

\section{\label{sec:pizza}Multiple-cell cavity}

The multiple cavity system is an array of smaller identical cavities fitting inside a magnet bore and is one of the most intuitive ways to minimize the detection volume loss when targeting high mass axions~\cite{UCD2001Kinion,MCD2020Yang}.
However, since there is always packing loss due to the cavity thickness and the shape of the cavities, it is difficult to make a full use of the given magnet bore.
Also, all the individual cavities should be finely tuned to the same resonant frequencies within the cavity bandwidth to effectively enhance the sensitivity.
This implies that frequency synchronization will be more difficult for a larger number of cavities or a higher cavity quality factor~\cite{AP2018Jeong}.
The receiver chain also becomes complicate because multiple antennae and an RF combiner are required to extract and combine the signal from individual cavities.

The multiple-cell cavity , on the other hand, was designed to overcome such weaknesses, improving the detection efficiency for high mass axion searches.
A multiple-cell cavity is constructed by vertically dividing a cylindrical cavity into pizza-slice-shaped cells using thin metal partitions at equidistant intervals~\cite{PLB2018Jeong}.
This design is distinct from superconducting RF cavities, routinely used in particle accelerators, which are also applicable for axion haloscope searches~\cite{ARXIV2022Giaccone}.
The insertion of thin partitions is a key idea of the design for minimizing the volume loss while increasing the resonant frequency.
The partitions alter the original field distribution of the TM$_{010}$ mode, eventually forming azimuthal nodes at the position of the partitions.
The configuration gives rise to higher resonant modes resembling TM$_{n10}$, where $n$ is half the number of the divided cells.
Those TM$_{n10}$-like modes are degenerated at a same frequency.
The degeneracy is broken by introducing a partition gap in the middle of the cavity, shown in the figure in Table~\ref{tab:multicell_comparison}.
The lowest resonant mode, TM$_{010}$-like, is our desired mode because the EM field of each cell oscillates in phase, yielding the maximal form factors under a static magnetic field.
Another important benefit from the partition gap is that all the cells become spatially interconnected, enabling the signal to be extracted from the entire cavity volume using a single antenna and thus eliminating the necessity of a signal combiner.
Such features considerably simplify the structure of the receiver chain and facilitates experimental design.
Table~\ref{tab:multicell_comparison} summarizes the pros and cons of the multiple cavity system and the multiple-cell cavity design.
How the latter is robust against fabrication tolerance is discussed in great detail in Sec.~\ref{sec:local}.

\begin{table}
\caption{\label{tab:multicell_comparison}Comparison between the multiple-cavity system and the multiple-cell cavity design. The thick brown lines represent the metallic cavity surface and the black dashed lines refer to the magnet bore.}
\centering
\begin{tabular}{lcc}
 \hline
        & Multiple cavity system & Multiple-cell cavity \\ \hline \hline
 Sketch & \parbox[c][0.205\linewidth]{0.2\linewidth}{\includegraphics[width=\linewidth]{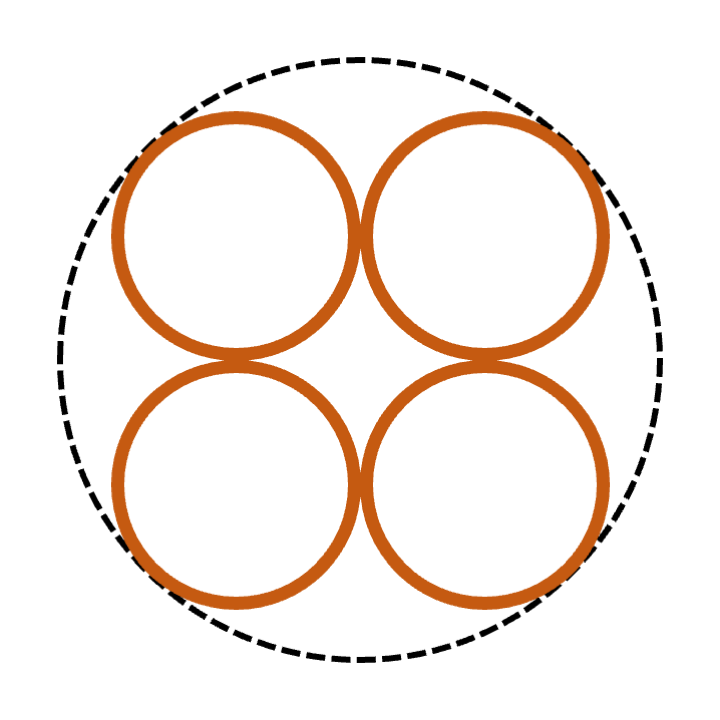}} & \parbox[c][0.205\linewidth]{0.2\linewidth}{\includegraphics[width=\linewidth]{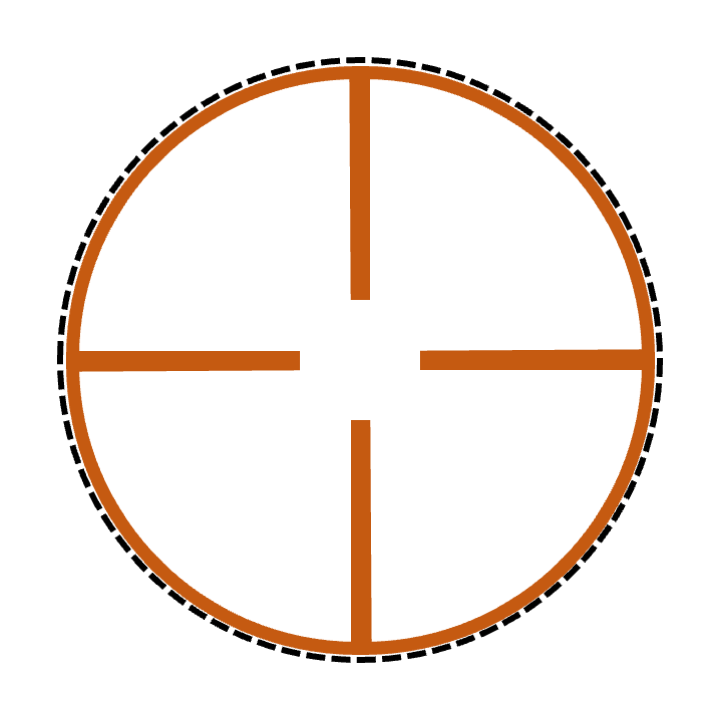}} \\ \hline
 \multirow{3}{*}{Pros} & & Minimal volume loss \\
 & Intuitive & Single antenna \\
 & & Robust against tolerance \\ \hline
 \multirow{3}{*}{Cons} & Inefficient in volume & \\
 & Multiple antennae & Difficult to fabricate \\
 & Frequency matching & \\ \hline
\end{tabular}
\end{table}

\section{\label{sec:chara}Characteristics}

The characteristics of multiple-cell cavities are determined by the cavity radius, the number of cells, and the size of the center gap.
For a small center gap, the radius and central angle have a leading effect on the field solution in each cell such that analytical calculations are possible to determine the resonance frequency, quality factor, and form factor.
However, the field solution significantly varies as the center gap becomes large, requiring numerical approaches to study the characteristics.

\begin{figure}
\centering
\includegraphics[width=0.8\linewidth]{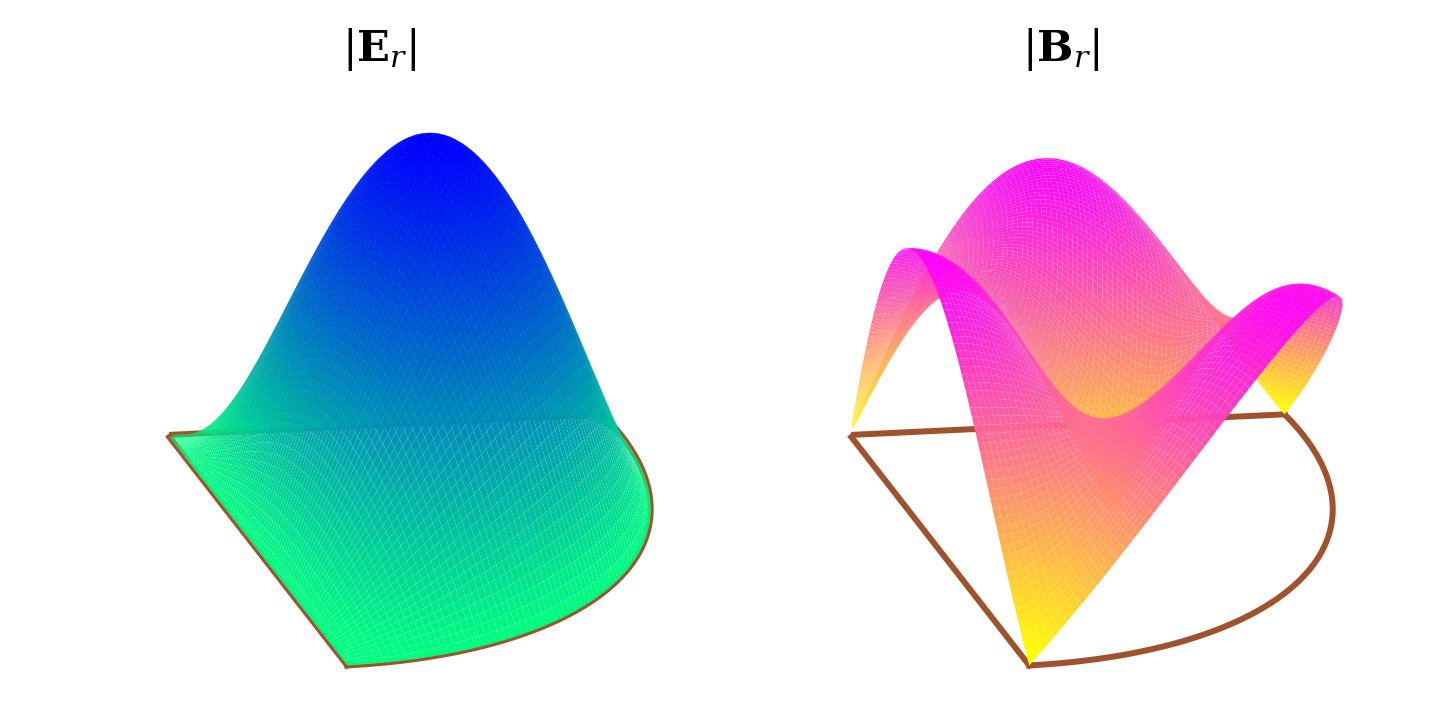}
\caption{Norm of the electromagnetic fields for the lowest TM mode of a circular sector cavity. The color and height indicate the field strength.} 
\label{fig:solution}
\end{figure}

For a closed circular sector with radius $R_c$, the electromagnetic field solution of the lowest TM mode is given by the Bessel functions as follows in cylindrical coordinates $\rho$, $\varphi$, and $z$:
\begin{equation}
\begin{split}
    \mathbf{E}_{r} &= E_{0} J_{m}\left( \chi_{m1}\frac{\rho}{R_{c}} \right) \cos(m\varphi) e^{-i \omega_{c} t} \hat{z}, \\
    \mathbf{B}_{r} &= \frac{1}{i\omega_{c}} \nabla \times \mathbf{E}_{r},
\end{split}
\end{equation}
where $E_{0}$ is the amplitude of the electric field, $J_{m}$ is the Bessel function of the first kind of the $m$-th order, and $\chi_{m1}$ is the first root of $J_{m}$.
The resonant frequency $\omega_{c} (=2\pi\nu_c)$ satisfies $\omega_{c} = c \chi_{m1} / R_{c}$ where $c$ is the speed of light.
The subscript $m$ represents the azimuth node number, determined by the angle of the sector $\phi$, which yields $m = \pi / \phi$ from the boundary condition of $\mathbf{E}_{r}(\varphi=\pm\phi/2) = 0$.
The root of the Bessel function increases with decreasing $\phi$, so the more a cavity is divided, the higher the resonant frequency.
An example of the field distributions for $m=4$ are shown in Fig.~\ref{fig:solution}.

The quality factor, defined as the ratio of the stored energy per resonance cycle to the dissipation power, can be exactly calculated from the field solution.
For a circular sector cavity with radius $R_{c}$, length $L_{c}$, and azimuth node number $m$, the quality factor is:
\begin{equation}
    Q_{c} = \frac{1}{1/\mathcal{Q} + R_{c} / L_{c}} \frac{R_{c}}{\delta_{c}},
\end{equation}
with
\begin{equation}
\label{eq:q_coeff}
\begin{split}
    \frac{1}{\mathcal{Q}} = \frac{1}{4\pi J_{m+1}(\chi_{m1})^{2}}\bigg[ & \pi\left(J_{m-1}(\chi_{m1})-J_{m+1}(\chi_{m1})\right)^{2} \\
    & + \frac{32 m^{3}}{4 m^{2}-1}J_{m+1}(\chi_{m1})^{2} \bigg],
\end{split}
\end{equation}
where $\delta_{c}$ is the skin depth of the cavity surface and $\mathcal{Q}$ implies the quality factor normalized to its radius and the skin depth assuming sufficiently long cavity length.

Likewise, the form factor is also exactly calculable from the field solution using its definition of Eq.~\ref{eq:form_factor}.
Assuming a homogeneous magnetic field, $\mathbf{B}_{0} = B_{0}\hat{z}$, the form factor can be expressed as 
\begin{equation}
\begin{split}
    C =& \frac{2^{3 - 2m}(\chi_{m1})^{2m}}{\pi^{2}J_{m+1}(\chi_{m1})^{2}} \left[\Gamma \left(\frac{m}{2} + 1 \right) \right]^{2} \\
    & \left[ {}_{1}\tilde{F}_{2}\left(m/2 + 1; m/2 + 1, m + 1; -\frac{\chi_{m1}^{2}}{4}\right) \right]^{2},
    \label{eq:form_cal}
\end{split}
\end{equation}
where $\Gamma$ is an Euler gamma function and ${}_{p}\tilde{F}_{q}$ is a regularized generalized hypergeometric function.

Table~\ref{tab:multicell_charac} summarizes the characteristic parameters of a circular sector cavity for various values of $m$.
The resonant frequency increases as the number of partitions increases, for example, 8 partitions increases the frequency by more than a factor of 3.
With increasing partition number, the surface-to-volume ratio increases and thus the quality factor decreases.
Compared to the quality factor of cylindrical cavities with the same resonant frequencies, $\mathcal{Q}_{\rm cyl.}$, the reduction is not substantial.
The form factor maintains its value above 0.6 up to 8 partitions.
Therefore, the multiple-cell cavity design is useful for accessing high-frequency regions without significant degradation of the key parameters of haloscope.

\begin{table}
\caption{\label{tab:multicell_charac}Characteristic parameters of a circular sector cavity with different values of $m$.
$\mathcal{Q}_{\rm cyl.}$ is the quality factor compared to that of the cylindrical cavity with the same resonant frequency.
The quantities, except the form factor, are normalized to those for $m=0$, which corresponds to a conventional cylindrical cavity.
$\Delta\nu_c/\nu_c$ represents the frequency tunability achievable using an optimal set of dielectric rods depicted in the text.
}
\centering
\begin{tabular}{c|ccccc}
\hline
$m$ & ~~~0~~~ & ~~~1~~~ & ~~~2~~~ & ~~~3~~~ & ~~~4~~~ \\
\# of partition & 0 & 2 & 4 & 6 & 8 \\
\hline \hline
$\nu_{c}{}$ & 1 & 1.59 & 2.14 & 2.65 & 3.16 \\
$\mathcal{Q}{}$ & 1 & 0.54 & 0.42 & 0.34 & 0.28 \\
$\mathcal{Q}_{\rm cyl.}$ & 1 & 0.86 & 0.91 & 0.89 & 0.88 \\
$C$ & 0.69 & 0.64 & 0.65 & 0.63 & 0.61 \\ 
\hdashline
$\Delta\nu_c/\nu_c$ & 0.17 & 0.17 & 0.17 & 0.16 & 0.16 \\ 
\hline
\end{tabular}
\end{table}

For frequency tuning, equivalently field variation, the same mechanism as for the conventional multiple cavity system can be used with a dielectric or conducting rod inserted into each cell~\cite{AP2018Jeong}.
Because the multiple-cell design has a discrete symmetry in the azimuthal direction, the resonant frequency can be tuned by moving the rods along the same direction.
This can be achieved by rotating the set of tuning rods simultaneously with respect to the center of the cavity.
The tuning path is determined to maximize the tunable frequency range such that a rod passes through the point where the highest electric field of the resonant mode is formed.
The tuning mechanism for the multiple-cell cavity design is described in more detail in Section~\ref{sec:tuning}.

\begin{figure}
\centering
\includegraphics[width=0.8\linewidth]{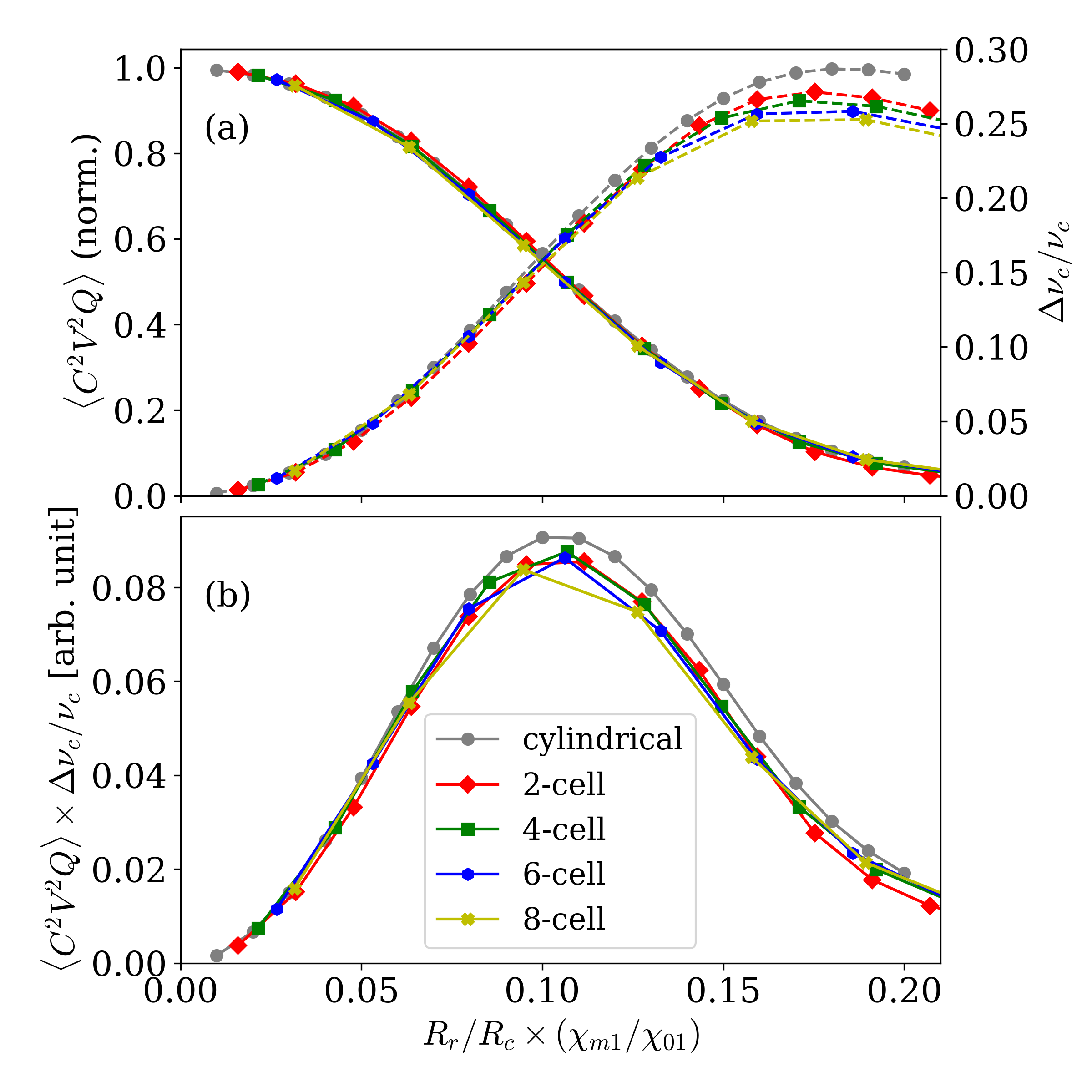}
\caption{
(a) Simulation results for scan rate (solid lines) and tuning range (dashed lines) as a function of rod radius relative to the cavity radius for different cell numbers.
The frequency dependence is taken into account by the increment factor $\chi_{m1} / \chi_{01}$ discussed in the text.
The dielectric constant is assumed to be 10.
(b) Product of the scan rate and tuning range.
} \label{fig:rod_radius}
\end{figure}

The size of the tuning rod affects both the tunable range $\Delta \nu$ and the average scan rate $\langle d\nu / dt \rangle$ ($\propto \langle C^{2}V^{2}Q \rangle$) -- in general, the greater the field variation by the rod, the larger the tunable range.
However, for dielectric rods, as the size increases, the form factor decreases and so does the average scan rate.
Therefore, it is necessary to optimize the rod dimension, typically its radius $R_r$, to obtain a sufficient scan rate over a reasonable tuning range.
Ref.~\cite{PATRAS2016Jeong} found that for a cylindrical cavity with a dielectric tuning rod with $\varepsilon=10$, $R_r/R_c\sim10\%$ maximizes the product of the average scan rate and the tunable range based on simulations.
A similar simulation study was repeated for various circular sectors and the results are shown in Fig.~\ref{fig:rod_radius}.
Since the wavelength of the mode varies with the number of cells, the length scale is normalized by the frequency increment factor, $\chi_{m1} / \chi_{01}$.
It is noticed from the figure that 1) the optimal rod size is consistent with that of a cylindrical cavity, and 2) the product of the average scan rate and the tunable range are approximately the same regardless of the cell multiplicity.
The relative tunable ranges are shown in the last row of Table~\ref{tab:multicell_charac}.

\section{\label{sec:tuning}Frequency tuning mechanism}
The basic concept of the frequency tuning system for the multiple($N$)-cell cavity design is described in Refs.~\cite{PLB2018Jeong, PRL2020Jeong}.
It consists of $N$ identical dielectric rods, one for each cell.
At the top and bottom of each cell, an arch-shaped opening is introduced through which the tuning rod can be extended out of the cavity.
The opening stretches from the partition to the field center of the cell such that the rod translates along the azimuthal direction with respect to the cavity center.
The set of tuning rods are gripped at both ends by a pair of holding structures outside the cavity, enabling all the rods to move simultaneously by rotating the structure.

\begin{figure}
    \centering
    \subfloat[\label{fig:photo_6-cell_1}]{\includegraphics[width=0.61\linewidth]{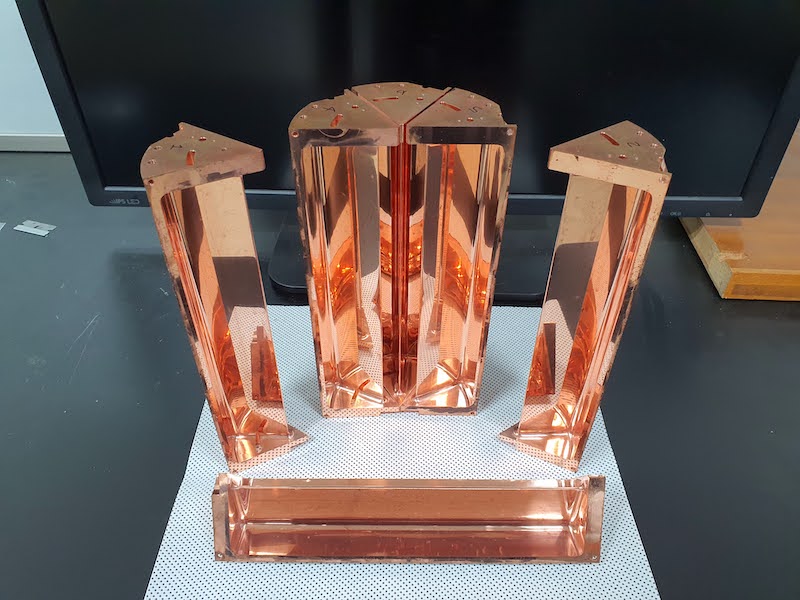}}
    \subfloat[\label{fig:photo_6-cell_2}]{\includegraphics[width=0.35\linewidth]{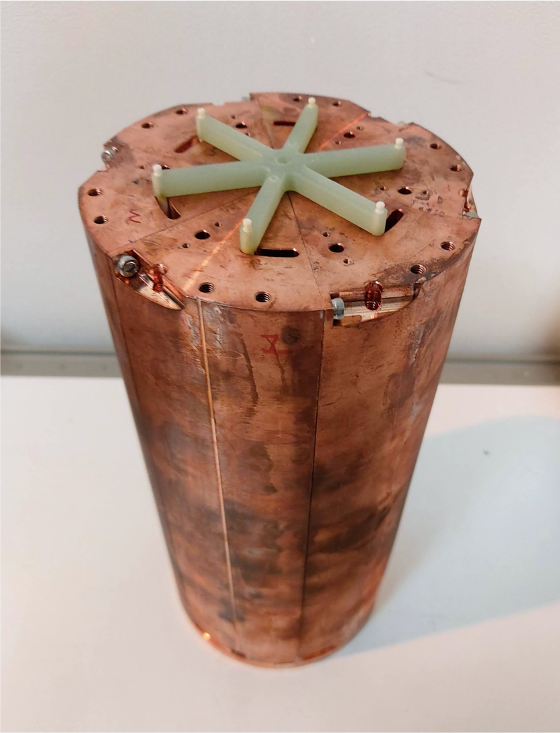}}
    \vfill
    \subfloat[\label{fig:Q_RT_6-cell}]{\includegraphics[width=0.8\linewidth]{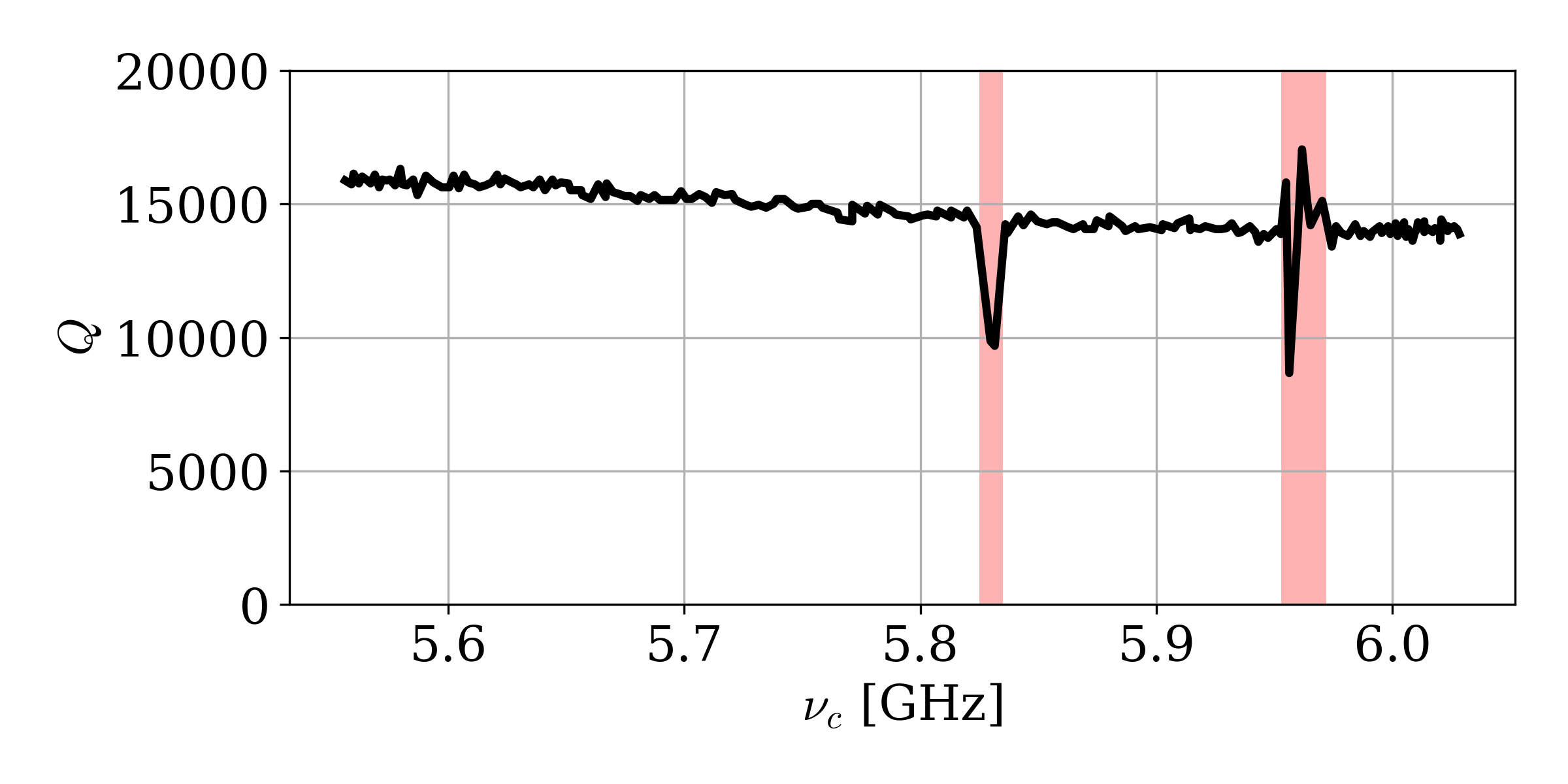}}
    \caption{(a) Photo of the individual cells of a 6-cell cavity.
    (b) Photo of the assembled caivity with a frequency tuning system installed.
    A set of alumina rods (white) are gripped at both ends by a star-shaped structure of G-10 (green) at the top and bottom of the cavity.
    (b) Quality factors measured in the process of frequency tuning at room temperature.
    The sudden drops, the red-painted regions, correspond to the mode crossings which are seen in the simulation.}
    \label{fig:6-cell}
\end{figure}

The tuning mechanism was demonstrated in Ref.~\cite{PLB2018Jeong} and applied in practice to an axion search in Ref.~\cite{PRL2020Jeong}, both using a double-cell cavity.
For a more general demonstration, we considered a multiple-cell cavity with $N=6$.
The cavity consisted of six identical copper pieces, each of which was fabricated as a single body with a 5mm-thick partition in the middle to form a 6-cell structure when assembled.
A photo of the individual cells is shown in Fig.~\ref{fig:photo_6-cell_1}.
The cavity dimensions were 110\,mm in inner diameter and 220\,mm in inner height.
A set of tuning rods were fabricated from 99.5\% aluminum oxide (Al$_2$O$_3$) with a diameter of 2.9\,mm.
Please note that a different rod dimension from the optimal value (3.8\,mm) was chosen for a dedicated axion search in the frequency range between approximately 5.5 to 6.0\,GHz.
The entire system was assembled as described above and a pair of star-shaped structures of G-10 were used to hold the tuning rods, as shown in Fig.~\ref{fig:photo_6-cell_2}.
A rotational piezoelectric actuator is mounted on one of the G-10 structures for simultaneous rotation of the tuning rods.
The structure also has a hole in the middle through which an antenna can be inserted into the cavity to pick up the signal and guide the rods for symmetric motion with respect to the cavity center as well.
Figure~\ref{fig:Q_RT_6-cell} demonstrates the tuning mechanism by showing a smooth behavior of the quality factors over the frequency tuning range measured at room temperature.

\section{\label{sec:local}Field localization}
\begin{figure}
\centering
\includegraphics[width=0.3\linewidth]{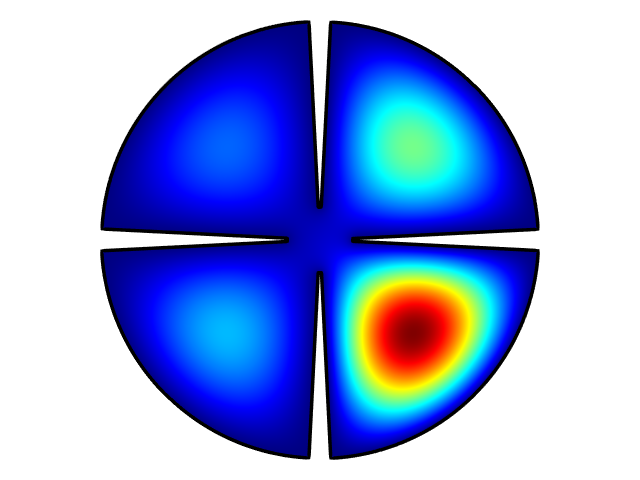}
\includegraphics[width=0.3\linewidth]{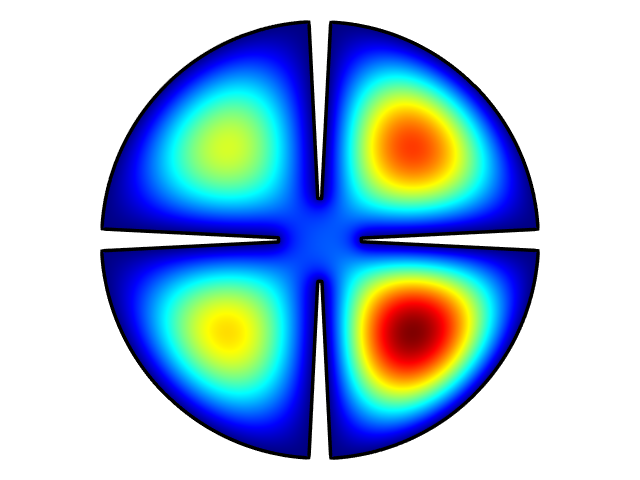}
\includegraphics[width=0.3\linewidth]{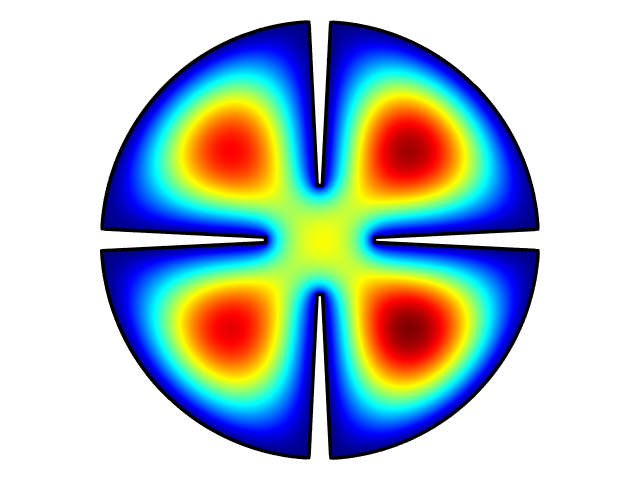}
\includegraphics[width=0.062\linewidth]{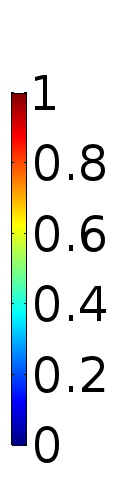}
\caption{Simulated electric field distributions for the TM$_{010}$-like mode of a 4-cell cavity with the dimensions altered by 0.05\%. The partition space increases from left to right.} 
\label{fig:localization}
\end{figure}

In the multiple-cell design, geometric asymmetry between cells will induce field localization to specific cells, which eventually degrades the form factor.
Naturally, the greater the asymmetry is, the greater the degree of localization will be.
A major cause of asymmetry is tolerances in cavity fabrication, including machining and assembly.
The tolerance effects can be estimated based on numerical simulation by slightly changing the dimensions of individual cells--the radius of each cell and the thickness and position of each partition.
Figure~\ref{fig:localization} shows some examples simulated using COMSOL Multiphysics software~\cite{COMSOL}.
Inspired by the fact that no considerable localization issue appears in cylindrical cavities, it is thought that for multiple-cell design, the degree of localization could be reduced by increasing the partition space in the middle of the cavity.
The effect is evaluated by repeating the above-mentioned simulation process with different values of partition gap.
Figure~\ref{fig:localization} illustrates how the field localization is being alleviated, i.e., the field becomes more evenly distributed over the cavity, with increasing center gap.

\begin{figure}
\centering
\includegraphics[width=0.8\linewidth]{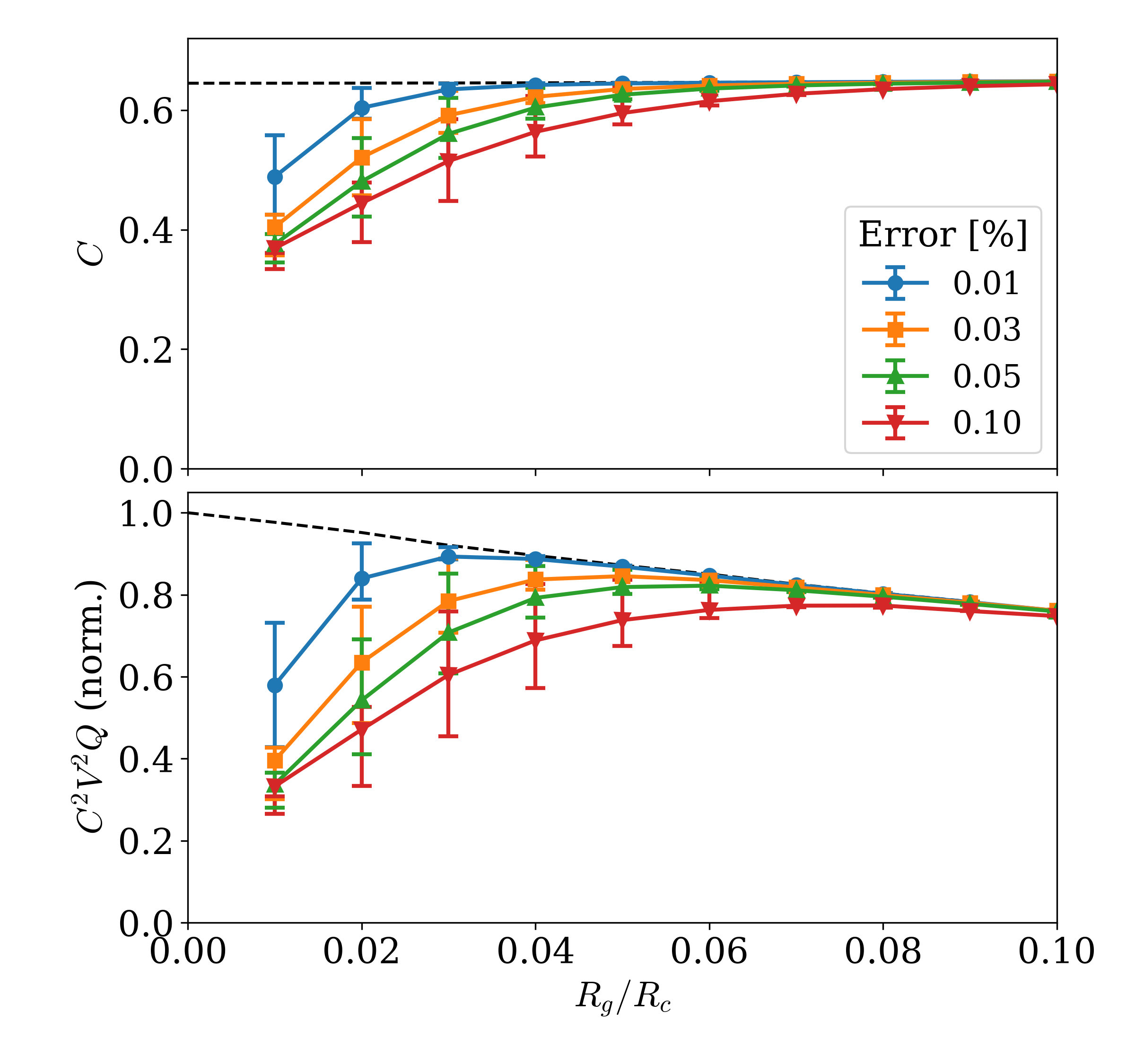}
\caption{Form factor $C$ and cavity-relevant quantity $V^2C^2Q$ for a 2-cell cavity as a function of partition gap for various tolerances.
The $V^2C^2Q$ values are normalized to those of the ideally modeled cavity.
Different colors represent different tolerance levels. The error bar at each point reflect the statistical deviation from the 100 simulations depicted in the text.
} \label{fig:2cell_tolerance_FOM}
\end{figure}

Increasing the partition gap effectively resolves the field localization issue. However, it draws more field into the center of the cavity, reducing the resonant frequency, losing the desired benefit for high-frequency searches.
Therefore, the center gap needs to be optimized to compromise these two aspects with fabrication tolerances taken into account.
Monte Carlo simulations using COMSOL were carried out for this purpose.
Beginning with an ideal cavity model, the cavity dimensions were allowed to randomly vary within 0.01\% to 0.10\% which are typical tolerances depicted in the international standard~\cite{ITgrade}.
Such perturbations were modeled 100 times and each time a quantity $V_c^2C^2Q_c$, a cavity-associated parameters in the scan rate Eq.~\ref{eq:scan_rate} to a measure of performance, was calculated.
This procedure was repeated for different sizes of partition gap ($R_g$), defined as the distance between the center of the cavity and the tip of a partition, to find the optimal value.
The results for a 2-cell cavity are shown in Fig.~\ref{fig:2cell_tolerance_FOM}, where the computed scan rates were normalized to that of a cavity without partition gap.
It is noticed that 1) a larger tolerance value degrades performance for a fixed gap, 2) the overall behavior converges to the ideal case with increasing gap size, and 3) for a larger tolerance level, the larger gap size is required to maximize the performance.
In general, precision fabrication will require smaller partition spacing to avoid field localization and thus achieve better performance.

\begin{figure}
\centering
\includegraphics[width=0.8\linewidth]{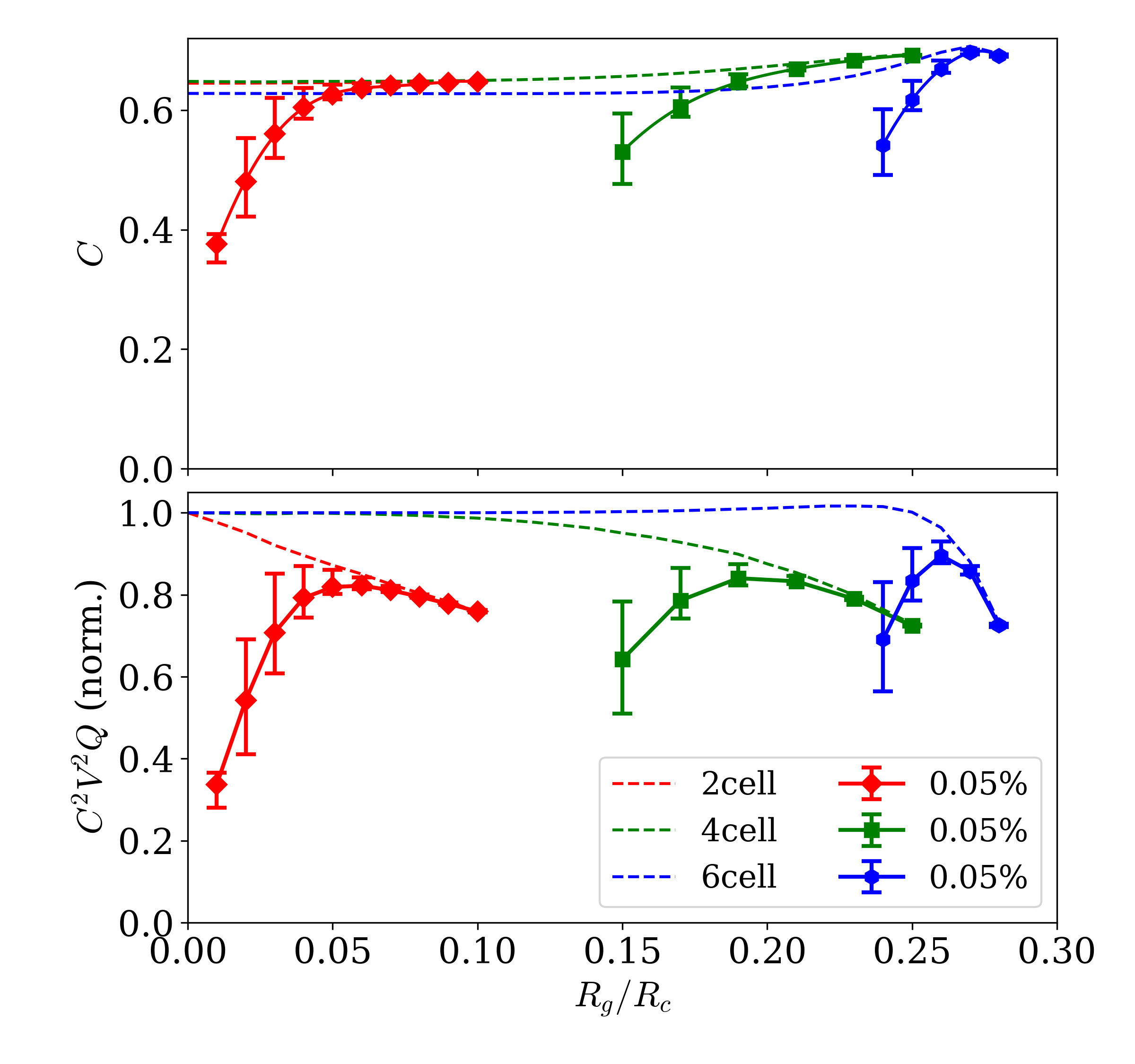}
\caption{Extended version of Fig.~\ref{fig:2cell_tolerance_FOM} to cavities with higher-cell multiplicities, 4-cell and 6-cell, represented in different colors. Only 0.05\% tolerance level was considered.} \label{fig:ncell_tolerance_FOM}
\end{figure}

The same simulation study was extended to cavities with higher-cell multiplicities.
Figure~\ref{fig:ncell_tolerance_FOM} shows the results for 4-cell and 6-cell cavities assuming 0.05\% tolerance level along with those for the 2-cell cavity.
It is found that higher-cell multiplicities require larger openings in the middle of the cavity for maximal performance.
It is also seen the scan rates relative to that of the ideal designs are comparable to each other regardless of cell multiplicity.
The numerical values of the optimal gap and the corresponding form factor and scan rate from Figs.~\ref{fig:2cell_tolerance_FOM} and~\ref{fig:ncell_tolerance_FOM} are summarized in Table~\ref{tab:optimal_cg}.

\begin{table}
\caption{\label{tab:optimal_cg}Optimal center gap and the corresponding form factor and scan rate for various multiple-cell cavities and tolerance levels.
The form factors were normalized to that for a 0.00\% tolerance level (i.e., ideal cavities), while the scan rates were normalized to that for $R_g=0$ (i.e., no gap).}
\centering
\begin{tabular}{cc|ccc}
\hline
~$m$~ & Error [\%] & ~$R_{g}^{\rm opt.}/R_{c}$~ & ~$C/C_{\rm ideal}$~ & ~$C^{2}V^{2}Q$ (norm.)~ \\
\hline \hline
\multirow{4}{*}{1} & 0.01 & 0.03 & 0.99$\pm 0.01$ & 0.90$\pm 0.01$ \\
% & 0.02 & 0.040 & 0.87 \\
 & 0.03 & 0.05 & 0.98$\pm 0.01$ & 0.85$\pm 0.02$ \\
% & 0.04 & 0.052 & 0.83 \\
 & 0.05 & 0.06 & 0.98$\pm 0.01$ & 0.82$\pm 0.02$ \\
 & 0.10 & 0.07 & 0.97$\pm 0.01$ & 0.78$\pm 0.01$ \\
2 & 0.05 & 0.20 & 0.98$\pm 0.01$ & 0.84$\pm 0.02$ \\
3 & 0.05 & 0.26 & 0.97$\pm 0.01$ & 0.90$\pm 0.02$ \\ \hline
\end{tabular}
\end{table}

\begin{figure}
    \centering
    \includegraphics[width=0.95\linewidth]{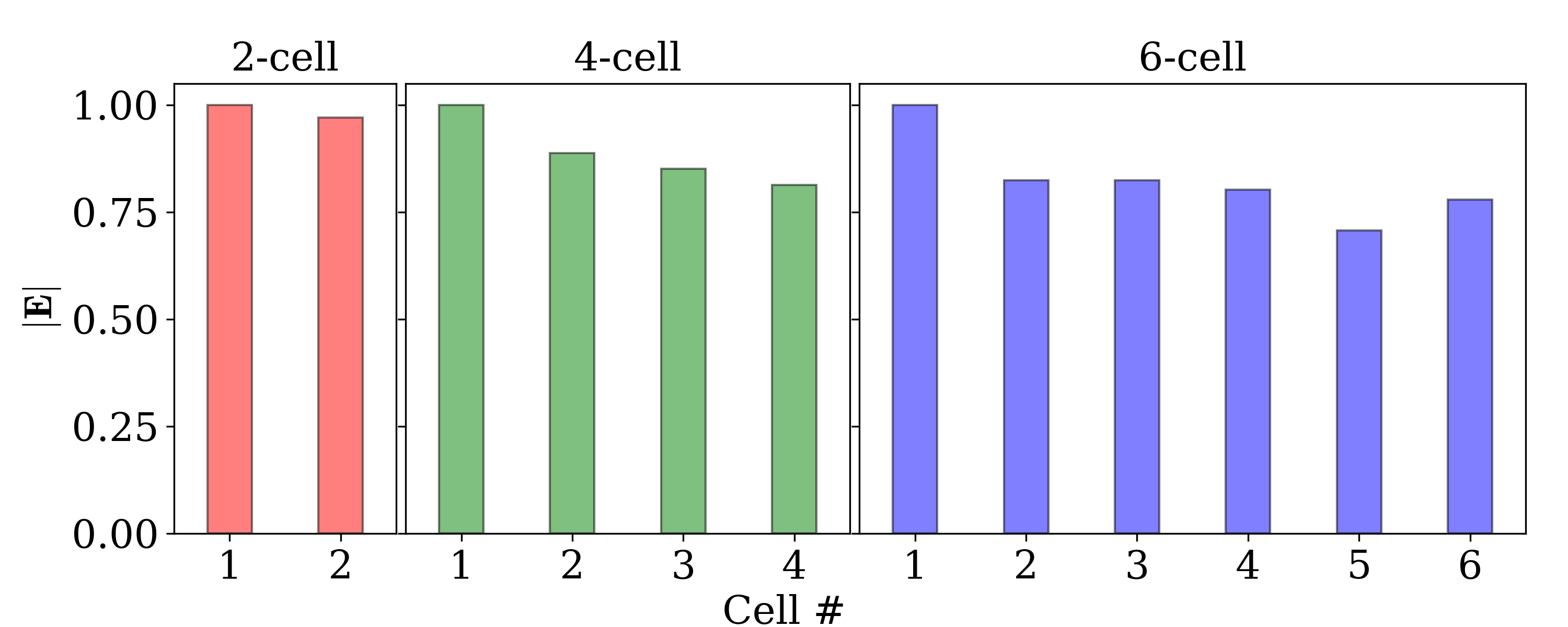}
    \caption{Distributions of electric field strength converted from the measured reflection signals of individual cells for the 2-, 4-, and 6-cell cavities.
    The cell with the maximum value is assigned to cell \#1 and the field strengths are normalized to that value.}
    \label{fig:field_strength_246cells}
\end{figure}

The simulation results regarding the field localization effects were verified using several multiple-cell (2-, 4-, and 6-cell) cavities of commercially available oxygen-free copper.
The cavity dimensions are 110\,mm in inner diameter and 215\,mm in inner height.
Each cavity consists of identical copper pieces, each of which was fabricated as a single body with a 5mm-thick partition in the middle to form a multiple-cell structure when assembled, as can be seen in Fig.~\ref{fig:photo_6-cell}. 
The fabrication tolerance level was requested to be 0.05\% and the corresponding optimal center gaps were chosen from Table~\ref{tab:optimal_cg}.
The tolerance level was validated by measuring the thickness of individual cells at many different places.
Each cell of the cavities has a hole in the same location at the top, through which the electric field can be probed using an RF antenna by measuring the reflected signal via a network analyzer.
For an ideal cavity, the overall field should be evenly distributed, so the field intensity (reflected signal strength) of the individual cells should be equal, otherwise it would not be.
In other words, a discrepancy in field intensity indicates field localization.
The reflected signals of all the cells were measured one by one using a same monopole antenna made from a commercially available RF connector with a flange around it and an extended inner conductor.
Based on the measurements, the field uniformity over the entire cavity was evaluated and the form factor degradation was calculated.
The estimated form factors relative to the ideal values for the 2-, 4-, and 6-cell cavities were 1.00, 0.99, and 0.98, respectively, which are consistent with the simulated values in Table~\ref{tab:optimal_cg}.
The distributions of measured field strengths are shown in Fig.~\ref{fig:field_strength_246cells}.

\begin{figure}
    \centering
    \includegraphics[width=0.95\linewidth]{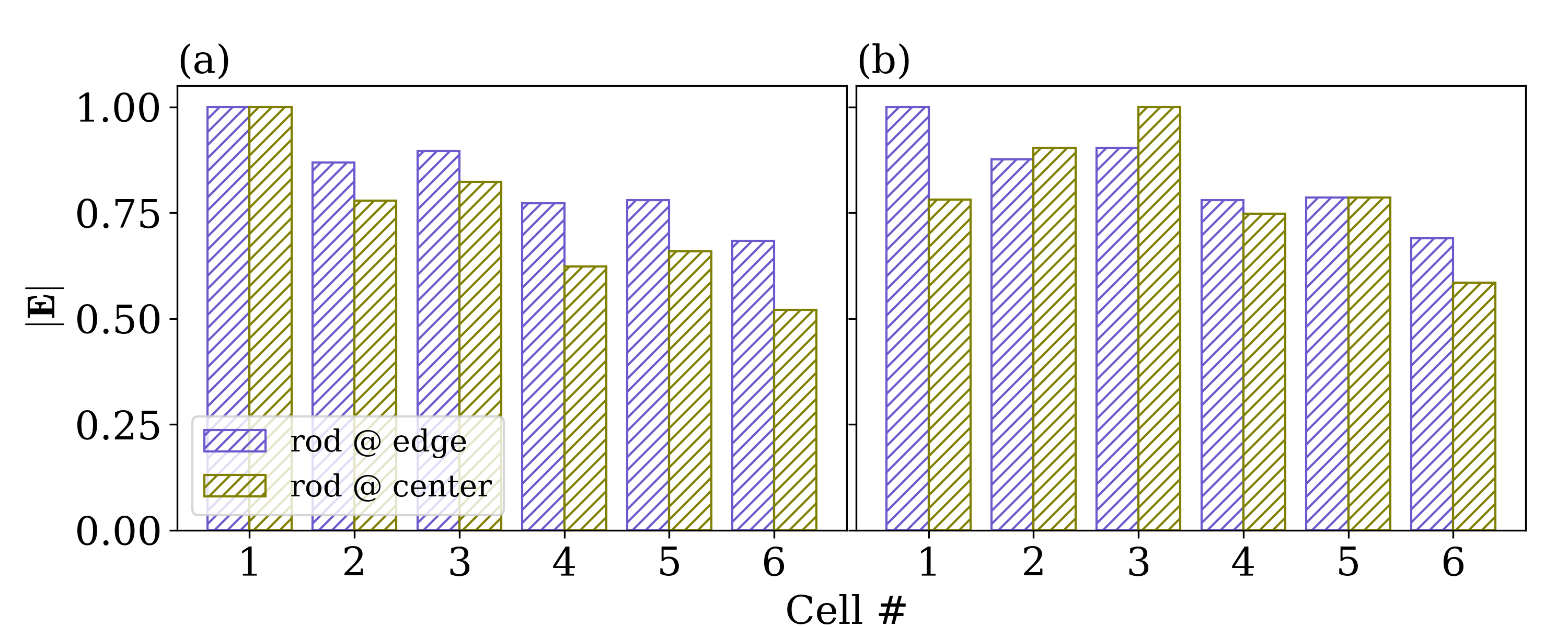}
    \caption{(a) Simulated field strength distributions of the 6-cell cavity, normalized to the maximum value, at two different rod positions. 
    (b) Distribution similar to (a) with the rod in cell \#1 replaced by one with the diameter reduced by 10\,$\mu$m.}
    \label{fig:field_strength}
\end{figure}

\begin{figure}
    \centering
    \includegraphics[width=0.95\linewidth]{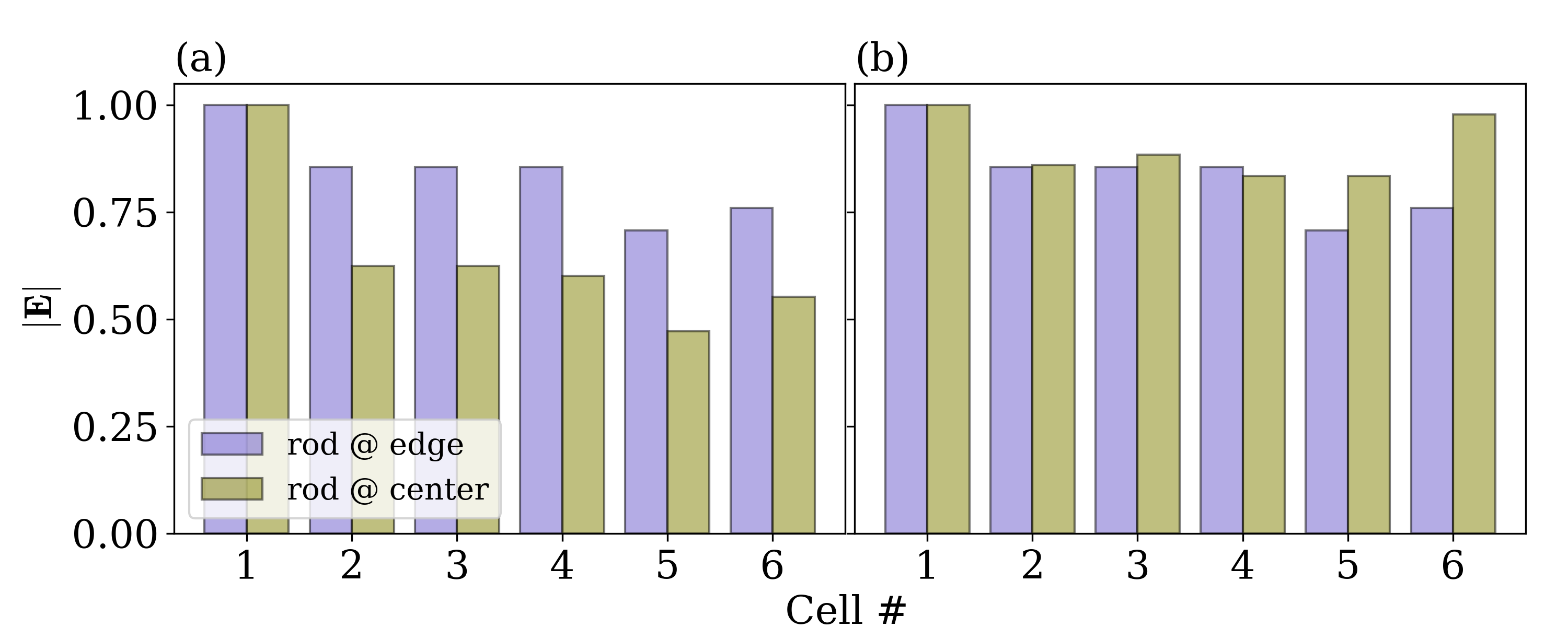}
    \caption{(a) Measured field strength distributions of the 6-cell cavity for different rod configurations. 
    (b) Distribution similar to (a) with the rod in cells \#1 and \#5 replaced by a 10\,$\mu$m-thinner and 10\,$\mu$m-thicker rod, respectively.}
    \label{fig:field_strength_6cell_rods}
\end{figure}

The field localization can also be attributed to the introduction of a geometrically asymmetric turning system inside the cavity.
The effect was evaluated from simulation by adding to each cell a dielectric rod whose position varies randomly within the same tolerance level.
We noticed more severe field localization, which was found to be due to the dielectric properties of attracting the electric field rather than due to the asymmetric geometry of the tuning system (less than 0.5\%).
The symptom was maximal (minimal) when the rod is positioned in the center (edge) of the field, yielding the relative form factor of $0.87\pm0.08$ $(0.97\pm0.03)$.
The simulation results were verified using the 6-cell cavity system, described in Section~\ref{sec:tuning}.
From the measurements of electric field strength of the individual cells, the estimated form factor was 0.94 (0.98) when the rods were placed in the center (edge) of the cells.
It was conceivable that such an effect could be mitigated by adjusting the dimension of the rods, e.g., reducing the diameter of the rod in the cell where the field is maximally confined in order to attract less field.
We implemented this idea in simulation by reducing the diameter of the corresponding rod by 10\,$\mu$m and observed an improvement in field uniformity by a few percent, which is shown in Fig.~\ref{fig:field_strength}.
For validation, we swapped two of the tuning rods in the 6-cell cavity system--one with a thinner rod in the cell having the highest field strength and the other with a thicker rod in the cell having the lowest field strength.
This configuration recovered the form factor by up to 5\%, i.e., 0.94 to 0.99 depending on the rod position.
The corresponding effects are illustrated in Fig.~\ref{fig:field_strength_6cell_rods}.

It should also be noted that the pick-up antenna is strongly coupled to the cavity through the hole in the top center in order to preserve the geometric symmetry. 
The effect of antenna misalignment was simulated by shifting the antenna position by 100\,$\mu$m from the center and found to be negligible.

\section{\label{sec:discussion}Discussion}

The multiple-cell cavity haloscope is expected to provide an efficient way to search for high mass dark matter axions.
The feasibility was successfully tested by performing a proof-of-concept experiment using a double-cell cavity at IBS-CAPP~\cite{PRL2020Jeong}.
This unique cavity concept has recently been adapted by various axion experiments.
KLASH proposed an experiment with multiple-cell cavities inside a large solenoid magnet to increase the search frequency~\cite{ARXIV2018KLASH}.
CAST-RADES also considered a similar concept for efficient axion searches within a long dipole magnet~\cite{JHEP2021RADES}.
In IBS-CAPP, a conventional cylindrical cavity of an existing experiment, whose the natural frequency is about 1.7\,GHz, was divided into eight cells to increase the target frequency around 6\,GHz.

Since the resonant frequency is determined by the cell number, i.e., the more cell number the high the resonant frequency, the multiple-cell haloscope, in principle, can be efficiently designed for any arbitrary search frequencies. 
This provides a useful feature particularly when multiple experiments with different detector sizes desire to aim the same search frequencies.
For example, IBS-CAPP plans to install 4-cell, 6-cell, and 8-cell cavities in the separate magnets of different dimensions, all of them aiming for near 6\,GHz in order to take advantage of phase-matched signal combination, which effectively enhance experimental sensitivity~\cite{AP2018Jeong}.

In addition to this cavity concept, there are other interesting ideas for accessing high-mass regions, such as exploiting higher-order resonant modes~\cite{JPG2020Kim,PRA2020Quiskamp} or employing the principle of metamaterials~\cite{PRL2019Lawson}.
These would offer a different approach independent of our multiple-cell concept.
This eventually indicates that if these additional ideas could be strategically implemented in each cell of a multiple-cell cavity, it would be possible to push the search domain into an even higher frequency region than the individual approaches could access.

\section{\label{sec:conclusion}Summary}

We performed an in-depth study on the multiple-cell cavity design to characterize it in a more general way using both analytical and numerical methods.
The major advantages over traditional cavity designs at high frequencies include 1) minimal volume loss, 2) simple experimental design, and 3) comparable cavity properties (form factor and quality factor).
The key experimental parameters remain high even with increasing cell multiplicity.
The potential field localization owing to machining tolerance is strategically mitigated by increasing the gap in the middle of the cavity.
The effect was confirmed by evaluating the field uniformity of several multiple-cell cavities.
%The experimental feasibility of this cavity concept was successfully demonstrated by employing a double-cell cavity into an axion search experiment at CAPP~\cite{PRL2020Jeong}.
This verifies that the multiple-cell haloscope can provide an efficient way to search for high-mass axions.

\section*{Acknowledgement}
J. Jeong and S. Youn are supported by the Institute for Basic Science (IBS-R017-D1-2022-a00).
J. E. Kim is supported by the National Research Foundation grants NRF-2018R1A2A3074631.

%  \bibliographystyle{elsarticle-num} 
%  \bibliography{cas-refs}

\begin{thebibliography}{00}

\bibitem{PRL1977PQ} R.D. Peccei and H.R. Quinn, Phys. Rev. Lett. {\bf 38}, 1440 (1977).
\bibitem{PRL1978Weinberg} S. Weinberg, Phys. Rev. Lett. {\bf 40}, 223 (1978).
\bibitem{PRL1978Wilczek} F. Wilczek, Phys. Rev. Lett. {\bf 40}, 279 (1978).
\bibitem{PLB1983Wilczek} J. Preskill, M.B. Wise and F. Wilczek, Phys. Lett. B {\bf 120}, 127 (1983).
\bibitem{PLB1983Abbott} L.F. Abbott and P. Sikivie, Phys. Lett. B {\bf 120}, 133 (1983).
\bibitem{PLB1983Dine} M. Dine and W. Fischler, Phys. Lett. B {\bf 120}, 137 (1983).
\bibitem{PRD1985Sikivie} P. Sikivie, Phys. Rev. D {\bf 32}, 2988 (1985).
\bibitem{JCAP2020Kim} D. Kim {\it et al.}, J. Cosmol. Astropart. Phys. {\bf 03}, 066 (2020).
\bibitem{PRL1979Kim} J. E. Kim, Phys. Rev. Lett. {\bf 43}, 103 (1979).
\bibitem{NPB1980SVZ} M.A. Shifman, A.I. Vainshtein and V.I. Zakharov, Nucl. Phys. B {\bf 166}, 4933 (1980).
\bibitem{YF1980Zhitnitsky} A.P. Zhitnitsky, Yad. Fiz. {\bf 31}, 497 (1980); Nucl. Phys. B {\bf 31}, 260 (1980).
\bibitem{PLB1981DFS} M. Dine, W. Fischler and M. Srednicki, Phys. Lett. B {\bf 104}, 199 (1981).
\bibitem{UCD2001Kinion} D.S. Kinion, PhD. Thesis, UC Davis (2001).
\bibitem{MCD2020Yang} J. Yang {\it et al.}, Microwave Cavities and Detectors for Axion Research {\bf 245} (2020).
\bibitem{AP2018Jeong} J. Jeong {\it et al.}, Astro. Phys. {\bf 97}, 33 (2018).
\bibitem{JPG2020Kim} J. Kim {\it et al.}, J. Phys. G: Nucl. Part. Phys. {\bf 47}, 035203 (2020).
\bibitem{PRA2020Quiskamp} A. P. Quiskamp {\it et al.}, Phys. Rev. Applied {\bf 14}, 044051 (2020).
\bibitem{PRL2017MADMAX} MADMAX Working Group, Phys. Rev. Lett. {\bf 118}, 091801 (2017).
\bibitem{PRL2019Lawson} M. Lawson {\it et al.}, Phys. Rev. Lett. {\bf 123}, 141802 (2019).
% \bibitem{NIM2021Alesini} D. Alesini {\it et al.}, Nucl. Instrum. Methods Phys. Res. A {\bf 985}, 164641 (2021)
\bibitem{PLB2018Jeong} J. Jeong {\it et al.}, Phys. Lett. B {\bf 777}, 412 (2018).
\bibitem{ARXIV2022Giaccone} B. Giaccone {\it et al.}, arXiv:2207.11346 (2022).
\bibitem{PRL2020Jeong} J. Jeong {\it et al.}, Phys. Rev. Lett. {\bf 125}, 221302 (2020).
\bibitem{PATRAS2016Jeong} J. Jeong, S. Youn,  S. Ahn and Y. K. Semertzidis, PATRAS {\bf 2016}, 48 (2017).
\bibitem{COMSOL} COMSOL Multiphysics$\textsuperscript{\textregistered}$ v. 5.2. www.comsol.com. COMSOL AB, Stockholm, Sweden.
\bibitem{ITgrade} International Organization for Standardization, ISO {\bf 286}, 1 (2010).
\bibitem{CIRP1991IKAWA} N. Ikawa {\it et al.}, CIRP Ann. Manuf. Technol. {\bf 40}, 2 (1991).
\bibitem{ARXIV2018KLASH} C. Gatti {\it et al.}, arXiv:1811.06754 (2018).
\bibitem{JHEP2021RADES} A. Á. Melcón {\it et al.}, J. High Energ. Phys. {\bf 2021}, 75 (2021).
\bibitem{PRL2020Lee} S. Lee, S. Ahn, J. Choi, B. R. Ko, and Y. K. Semertzidis, Phys. Rev. Lett. {\bf 124}, 101802 (2020).
\bibitem{PRL2021Kwon} O. Kwon, {\it et al.}, Phys. Rev. Lett. {\bf 126}, 191802 (2021).

\end{thebibliography}





% \end{thebibliography}

\end{document}